\documentclass[11pt,a4paper]{article}

\usepackage{epsfig}
\usepackage{amsmath}
\usepackage{amssymb}
\usepackage{verbatim}
\usepackage{bm}
\usepackage{dsfont}
\usepackage[footnotesize]{caption}
\usepackage{subfigure}
\usepackage{cancel}
\usepackage{cite}
\usepackage{textcomp}
\usepackage{calc}
\usepackage{geometry}
\usepackage{bbm}
\usepackage{color}

\begin{document}


\noindent July 2015
\hfill OSU-HEP-15-03

\hfill \parbox{\widthof{OSU-HEP-15-03}}{IPMU 15-0108}

\vskip 1.5cm

\begin{center}
{\LARGE\bf Pure Gravity Mediation and Spontaneous\\\smallskip
\boldmath{$B$$-$$L$} Breaking from Strong Dynamics}

\vskip 2cm

{\large K.~S.~Babu,$^{\negthinspace a}$
K.~Schmitz,$^{\negthinspace b}$ T.~T.~Yanagida\:$^{\negthinspace b}$}\\[3mm]
{\it{
a Department of Physics, Oklahoma State University, Stillwater, Oklahoma 74078, USA\\
b Kavli IPMU (WPI), UTIAS, The University of Tokyo, Kashiwa, Chiba 277-8583, Japan}}
\end{center}

\vskip 1cm


\begin{abstract}


In pure gravity mediation (PGM), the most minimal scheme for the mediation of
supersymmetry (SUSY) breaking to the visible sector, soft masses for the standard
model gauginos are generated at one loop rather than via direct couplings
to the SUSY-breaking field.
In any concrete implementation of PGM, the SUSY-breaking field is therefore
required to carry nonzero charge under some global or local symmetry.
As we point out in this note, a prime candidate for such a symmetry might be
$B$$-$$L$, the Abelian gauge symmetry associated with the difference between
baryon number $B$ and lepton number $L$.
The F-term of the SUSY-breaking field then not only breaks SUSY, but also $B$$-$$L$,
which relates the respective spontaneous breaking of SUSY and $B$$-$$L$
at a fundamental level.
As a particularly interesting consequence, we find that the heavy Majorana
neutrino mass scale ends up being tied to the gravitino mass, $\Lambda_N \sim m_{3/2}$.
Assuming nonthermal leptogenesis to be responsible for the generation of the baryon
asymmetry of the universe, this connection may then explain why SUSY
necessarily needs to be broken at a rather high energy scale, so that
$m_{3/2} \gtrsim 1000\,\textrm{TeV}$ in accord with the concept of PGM.
We illustrate our idea by means of a minimal model of dynamical SUSY
breaking, in which $B$$-$$L$ is identified as a weakly gauged flavor symmetry.
We also discuss the effect of the $B$$-$$L$ gauge dynamics on the superparticle mass
spectrum as well as the resulting constraints on the parameter space of our model.
In particular, we comment on the role of the $B$$-$$L$ D-term.


\end{abstract}


\thispagestyle{empty}

\newpage


\tableofcontents


\section{Introduction: SUSY and \boldmath{$B$$-$$L$} breaking by the same chiral field}
\label{sec:introduction}


Pure gravity mediation (PGM)~\cite{Ibe:2006de,Ibe:2012hu} is an attractive,
viable and minimal scheme for the mediation of supersymmetry (SUSY) breaking
to the visible sector.%
\footnote{For closely related mediation schemes, see~\cite{Hall:2011jd,Hall:2012zp}.}
The main idea behind this mediation scheme is that, given a rather high SUSY breaking
scale of $\mathcal{O}\left(10^{11}\cdots10^{12}\right)\,\textrm{GeV}$, soft SUSY breaking in
the minimal supersymmetric standard model (MSSM) can be solely achieved by means of
gravitational interactions.
In PGM, squarks and sleptons receive large masses of the order of the
gravitino mass, $m_{3/2} \sim 100\cdots 1000 \,\textrm{TeV}$, via
the tree-level scalar potential in supergravity (SUGRA)~\cite{Nilles:1983ge}.
Meanwhile, gauginos obtain one loop-suppressed masses around the TeV scale
via anomaly mediation (AMSB)~\cite{Dine:1992yw}.
Because of the large sfermion mass scale, PGM easily accounts for a standard model
(SM) Higgs boson mass of $126\,\textrm{GeV}$~\cite{Aad:2012tfa}, while, at the same time,
it is free of several notorious problems that other, low-scale realizations
of gravity mediation are usually plagued with (such as the cosmological gravitino
problem~\cite{Weinberg:1982zq} or the SUSY flavor problem~\cite{Gabbiani:1996hi}).


In particular, PGM does not suffer from the cosmological Polonyi
problem~\cite{Coughlan:1983ci}, which one typically encounters
in ordinary gravity mediation.
There, the SUSY-breaking (or ``Polonyi'') field  $X$ couples directly
to the chiral field strength superfields belonging to the SM gauge interactions,
\begin{align}
W \supset \frac{X}{M_{\rm Pl}} \mathcal{W}^\alpha \mathcal{W}_\alpha \,,
\label{eq:XWW}
\end{align}
with $M_{\rm Pl} = (8\pi G)^{-1/2} \simeq 2.44 \times 10^{18}\,\textrm{GeV}$
denoting the reduced Planck mass and which results in gaugino masses of
$\mathcal{O}\left(m_{3/2}\right)$.
To be able to write down such couplings in the superpotential,
one has to require that the field $X$ be completely neutral.
This, however, potentially leads to severe problems in the
context of cosmology.
Given a completely uncharged field $X$, the origin $X = 0$ does not have any
special meaning in field space, which is why $X$ is expected to acquire some vacuum
expectation value (VEV) of $\mathcal{O}\left(M_{\rm Pl}\right)$ during inflation,
$\left<X\right> \sim M_{\rm Pl}$.
In this case, a huge amount of energy ends up being stored in the coherent oscillations
of the Polonyi field after inflation.
Once released in the perturbative decay of the Polonyi field at late times, this
energy then results in dangerous entropy production as well as
unacceptably large changes to the predictions of big bang nucleosynthesis.
A number of solutions to this infamous Polonyi problem have been put forward over
the years in the context of ordinary gravity mediation (see, e.g., \cite{Linde:1996cx}).
At the same time, PGM resolves the Polonyi problem in the arguably simplest way,
i.e., by requiring that the origin of the Polonyi field
\textit{does} have a special meaning.
This is readily done by requiring SUSY to be broken by a \textit{non-singlet field},
i.e., in PGM, one assigns nonzero charge to the Polonyi field,
so as to single out the origin as a special point in field space.
During inflation, $X$ is then stabilized at $\left<X\right> = 0$,
due to a positive Hubble-induced mass around the origin, and we no longer have
to worry about large-amplitude oscillations of the Polonyi field after inflation.
This solves the Polonyi problem.
Meanwhile, given a charged SUSY-breaking field $X$, couplings between $X$
and the SM gauge fields such as in Eq.~\eqref{eq:XWW} are forbidden, which renders the
SM gauginos massless at tree level.
This is a characteristic feature (not a bug) of PGM, serving the purpose to lower
the gaugino masses relative to the sfermion mass scale by a loop factor down to the TeV
scale.
The lightest supersymmetric particle (LSP) may then correspond
to the wino, which may very well provide a viable candidate for dark matter (DM) in the form of
weakly interacting massive particles (WIMPs)~\cite{Ibe:2012hu,Ibe:2013jya,Hall:2012zp}.


A crucial question, which needs to be addressed in any implementation of PGM,
then is: Under which symmetry could the SUSY-breaking field $X$ be possibly charged?
Interesting candidates for such a symmetry are, e.g., a discrete $R$ symmetry,
a global or a local $U(1)$ symmetry.
Two of us have recently studied the first among these three scenarios
in more detail in~\cite{Harigaya:2015soa}, which is why we will not pay any
further attention to the possibility of a discrete $R$ symmetry in the following.
Similarly, the case of a global $U(1)$ symmetry has already been discussed in~\cite{Feldstein:2012bu}.
In this note, we shall therefore focus on the possibility of a local $U(1)$ symmetry
being responsible for vanishing gaugino masses at tree level. 
A prime candidate for such a protective $U(1)$ symmetry is $B$$-$$L$, the Abelian
gauge symmetry associated with the difference between baryon number $B$
and lepton number $L$.
This symmetry is essential to the seesaw mechanism~\cite{seesaw} and
may explain the origin of matter parity in the MSSM~\cite{Krauss:1988zc}.
In addition, it may also play an important role in the early universe
during the stages of reheating and leptogenesis (see, e.g., \cite{Buchmuller:2010yy}).
Furthermore, supposing that the field $X$ is indeed charged
under $B$$-$$L$, the auxiliary field $F_X$ also needs to carry
nonzero $B$$-$$L$ charge. 
In the SUSY-breaking vacuum at low energies, where $\left<\left|F_X\right|\right> \neq 0$,
the F-term of the SUSY-breaking field $X$ therefore not only breaks
SUSY, but also $B$$-$$L$.
Assuming, within the framework of PGM, that the gaugino mass
term in Eq.~\eqref{eq:XWW} is indeed forbidden by virtue of a local
$U(1)_{B-L}$, thus, establishes a link between the spontaneous
breaking of SUSY and the spontaneous breaking of $B$$-$$L$
at an elementary level.
As we shall argue in this paper, this has several interesting
phenomenological implications; most importantly, a direct connection
between the heavy neutrino mass scale $\Lambda_N$ in the seesaw extension of
the MSSM and the gravitino mass $m_{3/2}$,
\begin{align}
\Lambda_N \sim m_{3/2} \sim 100\cdots 1000 \,\textrm{TeV} \,.
\label{eq:LambdaNm32}
\end{align}
Note that this relation nicely embodies the connection between the spontaneous
breakings of $B$$-$$L$ and SUSY in our model, which is why it may be regarded
as the hallmark signature of our scenario.
As a consequence, the heavy Majorana neutrinos in the MSSM end up being much
lighter than usually expected according to, e.g., the standard
embedding of the seesaw mechanism into grand unified theories (GUTs).
Our scenario is, hence, inconsistent with the notion of standard
thermal leptogenesis (featuring a hierarchical heavy neutrino mass
spectrum)~\cite{Fukugita:1986hr}%
\footnote{Standard thermal leptogenesis requires the lightest sterile neutrino to have
a mass of at least $M_{N_1} \sim 10^9\,\textrm{GeV}$~\cite{Davidson:2002qv}.
Besides that, simple alternatives to the paradigm of thermal leptogenesis may easily involve
heavy Majorana neutrinos with masses almost as large as the scale of grand unification,
$M_{N_1} \sim 10^{15}\cdots10^{16}\,\textrm{GeV}$;
see~\cite{Kusenko:2014uta} for a recent example.\smallskip}
and, instead, requires some form of low-scale leptogenesis,
such as resonant leptogenesis~\cite{Pilaftsis:1997jf} (where the heavy neutrino masses
are highly degenerate), in order to account for the
baryon asymmetry of the universe.


The purpose of the present paper now is to illustrate our idea by means of a minimal
example.
More concretely, we shall demonstrate how to embed the spontaneous breaking of $B$$-$$L$
into one of the simplest models of dynamical SUSY breaking (DSB), i.e., the simplest realization of
the vector-like DSB model \`a la IYIT~\cite{Izawa:1996pk}, which is
based on strongly coupled $Sp(1) \cong SU(2)$ gauge dynamics in combination
with four fundamental matter fields.
Here, following up on earlier work presented in~\cite{Domcke:2014zqa},
we shall identify $B$$-$$L$ as a weakly gauged flavor symmetry of the IYIT model
(see Sec.~2).
Next to the anticipated link between the spontaneous breaking of SUSY and $B$$-$$L$
and the prediction for the heavy neutrino mass scale in Eq.~\eqref{eq:LambdaNm32},
this then provides us with important (partly tachyonic) corrections to the MSSM sfermion
masses.
These mass corrections consist, for one thing, of tree-level sfermion masses induced by
the $B$$-$$L$ D-term and, for another thing, of effective sfermion masses induced by
gauge mediation at the one-loop level~\cite{Intriligator:2010be} (see Sec.~3).
Both corrections need to be sufficiently suppressed
in order to ensure the stability of the low-energy vacuum.
Fortunately, as we shall discuss in more detail in Sec.~3, the suppression of the
$B$$-$$L$ D-term contributions to the MSSM sfermion masses turns out to be parametrically
well controlled, thanks to the fact that we are able to derive an \textit{explicit}
expression for the $B$$-$$L$ D-term in terms of the underlying model parameters.
In fact, owing to this calculability of the $B$$-$$L$ D-term, we are
capable of tuning its magnitude to an arbitrarily small value by
imposing an approximate flavor symmetry in the IYIT sector.
Our set-up therefore features an interesting mechanism to maintain control over
the $B$$-$$L$ D-term, which might otherwise spoil
large parts of our construction.%
\footnote{We believe that the applicability of this technical result may extend
well beyond the purposes of the present paper, which may make it also interesting
from a more general perspective, i.e., if one is more interested in
the general business of gauging global flavor symmetries of strongly coupled
DSB models and perhaps less interested in the concrete
phenomenology of a weakly gauged $B$$-$$L$ symmetry in the context of the IYIT model.}
Meanwhile, we find that the suppression of the gauge-mediated sfermion masses
imposes an upper bound on the $B$$-$$L$ gauge coupling constant, $g \lesssim 10^{-3}$,
which renders our model testable/falsifiable in a future multi-TeV collider experiment.
Finally, in Sec.~4, we are going to conclude,
giving a brief outlook as to how our study could possibly be continued.


\section{Embedding \boldmath{$B$$-$$L$} into the IYIT SUSY breaking model}


\subsection{Field content and low-energy effective theory}


In its most general formulation, the IYIT model of dynamical SUSY breaking is based on
a strongly coupled $Sp(N)$ gauge theory featuring $2N_f = 2(N+1)$
``quark'' fields $\Psi^i$ that transform in the fundamental representation of $Sp(N)$.
At energies below the dynamical scale $\Lambda$, this theory
is best described in terms of $(2N+1)(N+1)$ gauge-invariant composite ``meson'' operators
$M^{ij} \simeq \left<\Psi^i\Psi^j\right>/\left(\eta\Lambda\right)$, which are
subject to a quantum mechanically deformed moduli constraint~\cite{Seiberg:1994bz},
\begin{align}
\textrm{Pf}\left(M^{ij}\right) \simeq \left(\frac{\Lambda}{\eta}\right)^{N+1} \,, \quad
\eta \sim 4\pi \,.
\label{eq:constraint}
\end{align}
Here, $\textrm{Pf}\left(M^{ij}\right)$ denotes the Pfaffian of the antisymmetric meson matrix
$M^{ij}$ and $\eta$ is a numerical factor that may be estimated based on naive
dimensional analysis (NDA)~\cite{Manohar:1983md}.
In order to break SUSY in the IYIT model, one introduces Yukawa couplings
between the quark fields $\Psi^i$, the fundamental degrees of freedom (DOFs) at
energies above the dynamical scale $\Lambda$, and a set of $(2N+1)(N+1)$ singlet
fields $Z_{ij}$ in the tree-level superpotential,
\begin{align}
W_{\rm tree}^{\rm IYIT} = \frac{1}{2} \lambda_{ij}'\, Z_{ij}\, \Psi^i \Psi^j \,.
\label{eq:Wtree}
\end{align}
In the effective theory at energies below the dynamical scale $\Lambda$,
this gives rise to an effective superpotential for the meson fields $M^{ij}$, which lifts
all flat direction in moduli space,
\begin{align}
W_{\rm eff}^{\rm IYIT} \simeq \frac{1}{2}\lambda_{ij} \,\frac{\Lambda}{\eta}\,
Z_{ij}\, M^{ij} \,.
\end{align}
This superpotential implies F-term conditions for the singlet fields, $M^{ij} = 0$, 
which cannot be satisfied while simultaneously fulfilling the moduli constraint
in Eq.~\eqref{eq:constraint}, $\textrm{Pf}\left(M^{ij}\right) \neq 0$.
In the true vacuum of the IYIT model, SUSY is hence spontaneously broken
because some of the singlet fields' F-terms are nonzero, i.e., SUSY is broken
via the O'Raifeartaigh mechanism~\cite{O'Raifeartaigh:1975pr}.


For simplicity, we will restrict ourselves to the IYIT model in its simplest version
from now on.
That is, we will focus on the $Sp(1) \cong SU(2)$ case in combination with four
quark flavors.
For all Yukawa couplings in Eq.~\eqref{eq:Wtree} being equal,
$\lambda_{ij}' \equiv \lambda$,
the IYIT tree-level superpotential then exhibits a global $SU(4) \times Z_4$ flavor symmetry.%
\footnote{Here, the discrete $Z_4$ symmetry corresponds to a phase shift of all quark fields
by $\pi/2$, i.e., all quarks transform as $\Psi^i \rightarrow i\Psi^i$ under this $Z_4$
symmetry.
In fact, this $Z_4$ is nothing but the anomaly-free subgroup of the anomalous
$U(1)'$ symmetry that is contained in the full $U(4)$ flavor symmetry at the classical level,
$U(4) \cong SU(4) \times U(1)' \supset SU(4) \times Z_4$.}
Allowing for generic, numerically different couplings,
this symmetry is, however, broken down to an Abelian $U(1)_A \times Z_4$ flavor
symmetry.
In addition, the IYIT model always possesses a global, continuous and anomaly-free
$R$ symmetry, under which all quark fields carry charge $0$ and all singlet fields
carry charge $2$.
In summary, we therefore have
\begin{align}
\lambda_{ij}' \textrm{ all different} \quad\Rightarrow\quad
U(1)_R \times SU(4) \times Z_4 \:\:\rightarrow\:\: U(1)_R \times U(1)_A \times Z_4 \,,
\label{eq:breaking}
\end{align}
with the axial $U(1)_A \subset SU(4)$ being
associated with a global quark field rotation, $\Psi^i \rightarrow e^{iq_i\theta}\Psi^i$,
where all $U(1)_A$ charges $q_i$ sum to zero, $\sum_i q_i = 0$.
In \cite{Domcke:2014zqa}, this global $U(1)_A$ flavor symmetry has been promoted
to a weakly gauged Fayet-Iliopoulos (FI) symmetry, $U(1)_A\rightarrow U(1)_{\rm FI}$,
in order to demonstrate how to generate a theoretically consistent
and field-dependent FI D-term in the context of dynamical SUSY breaking.
The advantage of such dynamically generated FI-terms is that they do not
suffer from the usual problems that other FI models
are plagued with~\cite{Komargodski:2009pc,Dienes:2009td,Komargodski:2010rb}.
Once coupled to SUGRA, constant, field-independent FI-terms, e.g., always
require the presence of an \textit{exact} continuous global symmetry~\cite{Komargodski:2009pc},
which is problematic from the perspective of quantum gravity~\cite{Giddings:1988cx}.
On the other hand, field-dependent FI-terms in string theory~\cite{Dine:1987xk},
generated via the Green-Schwarz mechanism of anomaly cancellation~\cite{Green:1984sg},
imply the existence of a shift-symmetric modulus field~\cite{Komargodski:2010rb},
which causes cosmological problems~\cite{Coughlan:1983ci},
as long as it is not properly stabilized (which is hard~\cite{Binetruy:2004hh}).
As shown in \cite{Domcke:2014zqa}, dynamically generated and field-dependent
FI-terms in field theory, by contrast, avoid all of these problems, rendering
them the arguably best candidates for FI-terms with relevant implications for
low-energy phenomenology.


In this paper, we shall now take the analysis of \cite{Domcke:2014zqa} one step
further and promote the global $U(1)_A$ flavor symmetry of the IYIT model to a local
$U(1)_{B-L}$ symmetry.
If we assign $B$$-$$L$ charges $\pm q/2$ to the four fundamental quark fields at high energies,
the six composite meson fields at low energies end up carrying the following charges,
\begin{align}
\left[M_+\right] = +q \,, \quad \left[M_-\right] = -q \,, \quad 
\left[M_0^a\right] = 0 \,, \quad a = 1,2,3,4 \,,
\end{align}
and similarly for the six singlet fields $Z_\pm$ and $Z_0^a$, which we
also re-label according to their $B$$-$$L$ charges.
The effective superpotential as well as the effective K\"ahler potential
for these fields read%
\footnote{Throughout the analysis in this paper, we will take the K\"ahler
potential to be canonical for all fields and neglect all effects
induced by higher-dimensional terms in the effective K\"ahler potential.
These terms are uncalculable and, in principle, always present in the IYIT model.
On the other hand, they are suppressed compared to the canonical
terms in the K\"ahler potential by factors of
$\mathcal{O}\left(\lambda^2/\eta^2\right)$~\cite{Chacko:1998si},
which is why we can safely ignore them,
as long as we stay in the perturbative regime, $\lambda \ll \eta$, and do not
venture into the strongly coupled limit, where $\lambda \sim \eta$.}
\begin{align}
W_{\rm eff} \simeq & \: \frac{\Lambda}{\eta} \left(\lambda_+\, M_+\, Z_- +
\lambda_-\, M_-\, Z_+ + \lambda_0^a\, M_0^a\, Z_0^a\right) 
\label{eq:Weff} \,, \\ \label{eq:Keff}
K_{\rm eff} \simeq & \: M_+^\dagger e^{2qgV} M_+ + M_-^\dagger e^{-2qgV} M_-
+ Z_+^\dagger e^{2qgV} Z_+ + Z_-^\dagger e^{-2qgV} Z_- + 
\sum_a \left|M_0^a\right|^2 + \sum_a \left|Z_0^a\right|^2 \,.
\end{align}
Here, the vector field $V$ stands for the $B$$-$$L$ vector multiplet,
the auxiliary $D$ component of which
gives rise to the following D-term scalar potential,
\begin{align}
V_D = \frac{1}{2}D^2 = \frac{q^2g^2}{2} \left[\left|M_-\right|^2-\left|M_+\right|^2 +
\left|Z_-\right|^2-\left|Z_+\right|^2\right]^2 \,.
\end{align}


After imposing the quantum mechanically deformed moduli constraint in Eq.~\eqref{eq:constraint},
\begin{align}
\textrm{Pf}\left(M^{ij}\right) = M_+\, M_- - M_0^1\, M_0^4 + M_0^2\, M_0^3 \simeq
\left(\frac{\Lambda}{\eta}\right)^2 \,,
\end{align}
one finds that the vacuum manifold of the low-energy theory
exhibits exactly three local minima.
In the limit of a vanishingly small gauge coupling
constant $g$, these are respectively located at
\begin{align}
& \textrm{Vacuum I:} \quad & \left<M_+ M_-\right> \simeq & \left(\frac{\Lambda}{\eta}\right)^2 \,, &
\left<\left|M_+\right|\right> = & \:\frac{\sqrt{\lambda_+\lambda_-}}{\lambda_+} \frac{\Lambda}{\eta} \,, &
\left<\left|M_-\right|\right> = & \:\frac{\sqrt{\lambda_+\lambda_-}}{\lambda_-} \frac{\Lambda}{\eta} \,,
\label{eq:Vacua123}
\\\nonumber
& \textrm{Vacuum II:} \quad & - \left<M_0^1 M_0^4\right> \simeq & \left(\frac{\Lambda}{\eta}\right)^2 \,, &
\left<\left|M_0^1\right|\right> = & \:\frac{\sqrt{\lambda_0^1\lambda_0^4}}{\lambda_0^1} \frac{\Lambda}{\eta} \,, &
\left<\left|M_0^4\right|\right> = & \:\frac{\sqrt{\lambda_0^1\lambda_0^4}}{\lambda_0^4} \frac{\Lambda}{\eta} \,,
\\\nonumber
& \textrm{Vacuum III:} \quad & \left<M_0^2 M_0^3\right> \simeq & \left(\frac{\Lambda}{\eta}\right)^2 \,, &
\left<\left|M_0^2\right|\right> = & \:\frac{\sqrt{\lambda_0^2\lambda_0^3}}{\lambda_0^2} \frac{\Lambda}{\eta} \,, &
\left<\left|M_0^3\right|\right> = & \:\frac{\sqrt{\lambda_0^2\lambda_0^3}}{\lambda_0^3} \frac{\Lambda}{\eta} \,,
\end{align}
with all other meson and singlet VEVs vanishing, respectively.
The vacuum energies in these three vacua respectively scale with the
geometric means of the corresponding pairs of Yukawa couplings,
\begin{align}
V_{\rm I} = 2\,\lambda_+ \lambda_- \left(\frac{\Lambda}{\eta}\right)^4 \,, \quad 
V_{\rm II} = 2\,\lambda_0^1\lambda_0^4 \left(\frac{\Lambda}{\eta}\right)^4 \,, \quad
V_{\rm III} = 2\,\lambda_0^2\lambda_0^3 \left(\frac{\Lambda}{\eta}\right)^4 \,.
\label{eq:V123}
\end{align}
For $\lambda_+\lambda_- < \min\left\{\lambda_0^1\lambda_0^4,\lambda_0^2\lambda_0^3\right\}$,
the lowest lying vacuum therefore corresponds to the one where 
$\left<M_+ M_-\right> \simeq \left(\Lambda/\eta\right)^2$, i.e.,
the one in which $B$$-$$L$ is spontaneously broken by the
nonvanishing VEVs of the charged meson fields $M_\pm$.
In the following, we shall assume that this condition is satisfied, so that
in the low-energy vacuum of the IYIT model $B$$-$$L$
is indeed spontaneously broken.


In view of this result, two comments are in order:
(i) First of all, we remark that it is actually an open question whether the deformed
moduli constraint as stated in Eq.~\eqref{eq:constraint} really ends up being fulfilled
\textit{exactly} in the IYIT model or whether
$\textrm{Pf}\left(M^{ij}\right)$ could, in fact, also display a significant deviation
from $\left(\Lambda/\eta\right)^2$ in the true vacuum.
In the former case, it is only some of the singlet fields $Z_{ij}$ that acquire
nonzero F-terms, while in the latter case also the $Sp(N)$ glueball field
$T\propto \left<gg\right>$ turns out to contribute to SUSY
breaking with a nonzero F-term (see \cite{Domcke:2014zqa,Harigaya:2015soa}
for an extended discussion of this issue).
Our results will not be qualitatively affected by the choice between these 
two options, which is why, in this paper, we decide to neglect the possibility
of a dynamical glueball field and work with
$\textrm{Pf}\left(M^{ij}\right) \equiv \left(\Lambda /\eta\right)^2$
for simplicity in the following.
(ii) Our results in Eqs.~\eqref{eq:Vacua123} and \eqref{eq:V123} only
hold in the weakly gauged limit, $g\rightarrow0$. 
Once we turn on the $B$$-$$L$ gauge interactions, the vacuum manifold of
the IYIT model becomes distorted.
That is, while the loci of vacua II and III remain unchanged, vacuum I
begins to shift in the $M_\pm$ plane, as soon as the coupling $g$
is allowed to take a small, but nonzero value.
More precisely, for small $g$, we find
\begin{align}
\left<\left|M_\pm\right|\right> = 
\frac{\lambda}{\lambda_\pm} \frac{\Lambda}{\eta}
\left[1 \pm \frac{\gamma^2}{\rho^4}\left(1-\rho^4\right)^{1/2}
+ \mathcal{O}\left(\gamma^4\right)\right] \,,
\label{eq:Mpmgamma}
\end{align}
where we have introduced $\lambda$,  $\rho$ and $\gamma$ as important
combinations of the parameters $\lambda_\pm$ and $g$,
\begin{align}
\lambda = \sqrt{\lambda_+\lambda_-} \,, \quad
\rho = \left[\frac{1}{2}\left(\frac{\lambda_+}{\lambda_-} +
\frac{\lambda_-}{\lambda_+}\right)\right]^{-1/2} \,, \quad
\gamma = \frac{qg}{\lambda} \,.
\label{eq:lrg}
\end{align}
Here, $\lambda$ denotes the geometric mean of $\lambda_+$ and $\lambda_-$,
the parameter $\rho \in [0,1]$ is a convenient measure for the amount of
flavor symmetry violation in the charged meson sector,%
\footnote{Note that, for equal Yukawa couplings, $\lambda_+ = \lambda_-$, the
parameter $\rho$ goes to $\rho=1$, while, for drastically different Yukawa
couplings in the charged meson sector, $\lambda_+ \ll \lambda_-$ or
$\lambda_- \ll \lambda_+$, it approaches $\rho =0$.
Moreover, $\rho^2$ can also be interpreted as the ratio between the harmonic and
geometric means of $\lambda_+^2$ and $\lambda_-^2$ (see~\cite{Harigaya:2015soa}
for details).\label{fn:rho}}
and $\gamma$ characterizes
the strength of the $B$$-$$L$ gauge interactions relative to the strength of
the IYIT Yukawa interactions.
Eq.~\eqref{eq:Mpmgamma} illustrates that, while vacuum I always remains
on the $M_+M_- = \left(\Lambda/\eta\right)^2$ hypersurface, its
``flavor composition'' in terms of $M_+$ and $M_-$ begins to change in consequence
of the $B$$-$$L$ gauge interactions, once the gauge coupling strength $g$
takes larger and larger values.
A more detailed investigation of these next-to-leading order effects
in the gauge coupling constant $g$ is left for future work
(especially a study of the dynamics in the large-$g$ regime, where
$\gamma \gg 1$).
In this paper, we will, by contrast, content ourselves with a leading-order
analysis, meaning that wherever possible we will simply neglect
all effects of $\mathcal{O}(g)$.


\subsection{Particle spectrum in the vacuum at low energies}


So far, we have identified the condition under which the low-energy vacuum
of the IYIT model not only breaks SUSY, but also $B$$-$$L$.
Next, let us discuss the properties of this vacuum in a bit more detail.
In doing so, we will mostly review some earlier results presented
in~\cite{Domcke:2014zqa}, which is why we will be rather brief in what follows.
The physical mass eigenstates at low energies are contained in the following
two linear combinations of the singlet fields $Z_+$ and $Z_-$,
\begin{align}
X = \frac{1}{\sqrt{2}}\left(Z_+ + Z_-\right) \,, \quad 
Y = \frac{1}{\sqrt{2}}\left(Z_+ - Z_-\right) \,,
\label{eq:ZXY}
\end{align}
the $B$$-$$L$ vector multiplet $V$ as well as in the goldstone multiplet $A$
of spontaneous $B$$-$$L$ breaking,%
\footnote{In~\cite{Harigaya:2015soa} (making use of some
earlier results presented in \cite{Harigaya:2013vja}), the $U(1)_A$ flavor
symmetry of the IYIT model has been identified with the global Peccei-Quinn
(PQ) symmetry appearing in the axion solution to the strong $CP$ problem,
$U(1)_A \rightarrow U(1)_{\rm PQ}$, rather than with a local $B$$-$$L$
symmetry.
In this case, the field $A$ then turns out to correspond to the chiral axion
superfield in a supersymmetric version of the KSVZ axion model~\cite{Kim:1979if}.}
\begin{align}
M_\pm = \left<\left|M_\pm\right|\right> e^{\pm A/f_A} \,, \quad
f_A = K_0^{1/2} \,, \quad
K_0 = \big<\left|M_+\right|^2\big> + \big<\left|M_-\right|^2\big> \,.
\label{eq:MpmA}
\end{align}
Here, the decay constant $f_A$ ensures the correct normalization of the goldstone
field $A$ and $K_0$ represents the VEV of the K\"ahler potential in global SUSY.
While the actual goldstone phase $a\in A$ remains massless and is absorbed by the
$B$$-$$L$ vector field $A_\mu \in V$ upon spontaneous $B$$-$$L$ breaking, all other
DOFs contained in $A$ obtain soft SUSY-breaking masses via the IYIT superpotential.
The goldstone field $A$, hence, vanishes in the true vacuum, $\left<A\right> = 0$,
which allows us to expand the effective superpotential for $X$, $Y$ and $A$ in powers
of $A$.
Up to $\mathcal{O}\left(A^2\right)$, we have
\begin{align}
W_{\rm eff} \simeq \mu^2 X - m\, Y A + \frac{m^2}{2\,\mu^2}\, X A^2 \,, 
\label{eq:WXYA}
\end{align}
where $\mu$ and $m$ denote the F-term SUSY breaking scale as well as the soft
SUSY-breaking mass resulting from the IYIT superpotential, respectively,
(see Eq.~\eqref{eq:lrg} for the definitions of $\lambda$ and $\rho$)
\begin{align}
\mu = 2^{1/4} \lambda^{1/2} \, \frac{\Lambda}{\eta} \,, \quad
m = \frac{\mu^2}{f_A} = \rho\,\lambda\,\frac{\Lambda}{\eta} \,.
\label{eq:mum}
\end{align}
Correspondingly, the gravitino mass $m_{3/2}$ needs to take the following value
in our set-up,
\begin{align}
m_{3/2} = \frac{\mu^2}{\sqrt{3}\,M_{\rm Pl}} = 
\lambda \,\left(\frac{2}{3}\right)^{1/2} \frac{\left(\Lambda/\eta\right)^2}{M_{\rm Pl}} \,,
\label{eq:m32}
\end{align}
in order to ensure that the cosmological constant (almost) vanishes in the low-energy
vacuum.
Requiring the gravitino mass to take a certain value, say, $m_{3/2} = 1000\,\textrm{TeV}$,
thus allows us to eliminate either $\lambda$ or the dynamical scale $\Lambda/\eta$ from
our analysis.
We opt for the latter, so that
\begin{align}
\frac{\Lambda}{\eta} \simeq 1.7 \times 10^{12}\,\textrm{GeV}
\left(\frac{1}{\lambda}\right)^{1/2}
\bigg(\frac{m_{3/2}}{1000\,\textrm{TeV}}\bigg)^{1/2} \,.
\label{eq:Lambdaeta}
\end{align}


As evident from Eq.~\eqref{eq:WXYA}, $X$ corresponds to the SUSY-breaking goldstino
(or Polonyi) field.
Meanwhile, $Y$ pairs up with the chiral $B$$-$$L$ goldstone multiplet $A$ in
the superpotential in a such way that the fermionic components of $Y$ and $A$
share a Dirac mass term.
In terms of the charged meson fields $M_\pm$, the $F$ component of the goldstino
field $X$ is given as (see Eqs.~\eqref{eq:Weff} and \eqref{eq:ZXY})
\begin{align}
-F_X^* = \frac{1}{\sqrt{2}}\left(\lambda_+ M_+ + \lambda_- M_-\right)
\frac{\Lambda}{\eta} \,,
\end{align}
which acquires a VEV $\left<\left|F_X\right|\right> = \mu^2$ in the true vacuum.
Since $F_X$ does \textit{not} transform as a singlet under $B$$-$$L$, its nonzero
VEV not only breaks SUSY, but also $B$$-$$L$. 
We emphasize that this is one of the key features of the set-up
considered in this paper.
Furthermore, we note that $X$ is massless at tree level (see Eq.~\eqref{eq:WXYA}). 
At the classical level, the complex scalar contained in $X$, hence,
corresponds to a flat (or modulus) direction of the scalar potential.
This vacuum degeneracy is, however, lifted at the loop level~\cite{Chacko:1998si},
which renders the ``sgoldstino'' a pseudomodulus, after all.
The effective sgoldstino mass $m_X$ has recently been re-evaluated
in~\cite{Harigaya:2015soa} (see Eq.~(116) therein).
As it turns out, $m_X$ ends up being a complicated function of the Yukawa
couplings $\lambda_\pm$ and $\lambda_0^a$.
For this reason, we will not state the full expression here, but merely restrict
ourselves to the result in the flavor-symmetric limit, in which
$\lambda_0^a \equiv \lambda$ for all $a = 1,2,3,4$,%
\footnote{In this limit, the vacua I, II and III become degenerate
(see Eq.~\eqref{eq:V123}).
The breaking of the $SU(4)\cong SO(6)$ flavor symmetry down to $SO(5)$
then results in five massless particles: the $B$$-$$L$ goldstone phase
plus four genuine goldstone bosons, which may cause trouble at low energies.
Therefore, in order to avoid such massless particles, the global $SU(4)$ symmetry
should actually never be \textit{exactly} realized.
Instead, it should be at most realized as an \textit{approximate} symmetry,
so that all Yukawa couplings merely end up taking values \textit{close} to each other.
\label{fn:flavsymm}}
\begin{align}
m_X^2 = \frac{2\ln2-1}{16\pi^2}\left(1 + \frac{4}{\rho^6}\right)
\left(\frac{m}{\mu}\right)^4 m^2 \,.
\end{align}


Last but not least, it is instructive to examine the effective K\"ahler potential
for the charged meson fields $M_\pm$ as a function of $V$ and $A$
(see Eq.~\eqref{eq:Keff}).
Again expanding in powers of $A$, we find
\begin{align}
K_{\rm eff} = K_0 - 2 q g\, \xi\, V_A + m_V^2 V_A^2  + \mathcal{O}\left(V_A^3\right)
\,, \quad V_A = V + \frac{1}{\sqrt{2}\,m_V}\big(A + A^\dagger\big)
\,, \label{eq:KVA}
\end{align}
where $\xi$ denotes the $B$$-$$L$ FI parameter, $\xi \equiv \left<D\right>/(qg)$,
and $m_V$ is the $B$$-$$L$ vector boson mass,
\begin{align}
\xi = \big<\left|M_-\right|^2\big> - \big<\left|M_+\right|^2\big> \,, \quad
m_V = \sqrt{2} q g f_A \,.
\label{eq:ximV}
\end{align}
Eq.~\eqref{eq:KVA} nicely illustrates how the goldstone field $A$
is eaten by the $B$$-$$L$ vector multiplet $V$ upon spontaneous $B$$-$$L$
breaking.
In terms of the parameters of our model, $\xi$ and $f_A$ are given as
\begin{align}
\xi =  \left(\frac{\lambda_+}{\lambda_-} - \frac{\lambda_-}{\lambda_+}\right)
\left(\frac{\Lambda}{\eta}\right)^2
= \frac{2\left(1-\rho^4\right)^{1/2}}{\rho^2}\left(\frac{\Lambda}{\eta}\right)^2 \,, \quad
f_A =  \left(\frac{\lambda_+}{\lambda_-}+\frac{\lambda_-}{\lambda_+}
\right)^{1/2}\frac{\Lambda}{\eta} = \frac{\sqrt{2}}{\rho}\frac{\Lambda}{\eta} \,.
\label{eq:xiFA}
\end{align}


\section{Phenomenological consequences for neutrinos and sparticles}
\label{sec:pheno}


\subsection{Connection between the heavy neutrino mass scale and the gravitino mass}


In the previous section, we have shown how the spontaneous breaking of $B$$-$$L$
may be accommodated in the IYIT model of dynamical SUSY breaking.
Let us now study the phenomenological implications of this embedding.
First of all, we note that our set-up offers an intriguing possibility
to generate Majorana masses for the right-handed neutrinos in the
seesaw extension of the MSSM.
Suppose that the charge $q$ of the meson fields $M_\pm$ is actually given
as $q = 2$.
Then, gravitational interactions at the Planck scale will result in the
following operators in the effective theory (above and below the dynamical
scale $\Lambda$, respectively),
\begin{align}
W \supset \frac{1}{2}\frac{c_i}{M_{\rm Pl}} \Psi^3 \Psi^4\,N_i N_i \,, \quad
W_{\rm eff} \supset \frac{1}{2}\frac{c_i}{M_{\rm Pl}} \frac{\Lambda}{\eta}\, M_-\,N_i N_i \,,
\label{eq:MmNN}
\end{align}
where the $c_i$ are dimensionless coefficients of $\mathcal{O}(1)$ and with the fields
$N_i$ denoting the left-handed superfields the fermionic components of which correspond
to the hermitian conjugates of the right-handed neutrinos needed for the seesaw mechanism.
Upon spontaneous $B$$-$$L$ breaking, these couplings then turn into Majorana
mass terms for the neutrino fields $N_i$,
\begin{align}
W_{\rm} \supset \frac{1}{2} M_i\, N_i N_i \,, \quad
M_i = c_i\, \Lambda_N \,, \quad \Lambda_N = \frac{1}{\rho}
\left[1\pm\left(1-\rho^4\right)^{1/2}\right]^{1/2}
\frac{\left(\Lambda/\eta\right)^2}{M_{\rm Pl}} \,,
\end{align}
where the sign in the square brackets depends on whether $\lambda_-$ is smaller
($+$) or larger ($-$) than $\lambda_+$.
The heavy neutrino mass scale therefore turns out to be tied
to the gravitino mass (see Eq.~\eqref{eq:m32})!
\begin{align}
\Lambda_N = \frac{\left(3/2\right)^{1/2}}{\rho\,\lambda}
\left[1\pm\left(1-\rho^4\right)^{1/2}\right]^{1/2} m_{3/2} \,.
\end{align}
In the flavor-symmetric limit, $\rho\rightarrow1$, we find in particular,
\begin{align}
\Lambda_N \simeq 1200 \,\textrm{TeV}
\left(\frac{1}{\lambda}\right)
\bigg(\frac{m_{3/2}}{1000\,\textrm{TeV}}\bigg) \,.
\end{align}
We emphasize that this relation between the heavy neutrino mass scale $\Lambda_N$
and the gravitino mass $m_{3/2}$ is one of the most important phenomenological
consequences of our model.


Next, before turning to the phenomenological implications of our model for the MSSM sparticle
spectrum, we mention in passing that a coupling of the neutrino fields $N_i$ to the singlet
field $Z_-$ would, by contrast, \textit{not} allow for a successful generation
of the heavy neutrino mass scale $\Lambda_N$.
In SUGRA, the field $Z_-$ acquires a VEV of
$\mathcal{O}\left(m_{3/2}\right)$~\cite{Domcke:2014zqa,Harigaya:2015soa},
which is why one might naively think that a coupling of the form $Z_-NN$
in the superpotential may also result in neutrino masses of
$\mathcal{O}\left(m_{3/2}\right)$.
This is, however, not so because of the large F-term of the field $Z_-$,
which results in additional mass terms for the scalar neutrino fields of
$\mathcal{O}\left(\mu\right)$. 
After diagonalizing the sneutrino mass matrix, one then finds that some of the
sneutrinos end up being tachyonic with masses of $\mathcal{O}\left(-\mu\right)$,
which renders the coupling $Z_-NN$ unfeasible.
Moreover, in presence of the operator $Z_-NN$, the neutrino fields could
easily become unstable and absorb the SUSY-breaking F-term of the Polonyi field
in their VEV, $\left<NN\right> \sim - \mu^2$, which would restore SUSY at
low energies.
Therefore, we actually have to make sure that the coupling $Z_-NN$ is forbidden,
since it will otherwise interfere with our mechanism to generate the mass scale
$\Lambda_N$ or even spoil our entire SUSY breaking model.
This is best done by invoking $R$ symmetry, under which the neutrino fields carry
charge $1$, the meson fields charge $0$ and the singlet fields charge $2$
(see our discussion related to Eq.~\eqref{eq:breaking}
as well as the comments on $R$ symmetry in the IYIT model and the MSSM in
\cite{Domcke:2014zqa,Harigaya:2013vja,Harigaya:2015soa}).
$R$ symmetry then allows the couplings in Eq.~\eqref{eq:MmNN}, but forbids
couplings of the form $Z_-NN$.


\subsection{Tree-level corrections to the MSSM sfermion masses}


A second important consequence of our set-up for low-energy phenomenology
are tree-level as well as loop-induced corrections to the masses
of the MSSM sfermions.
Here, the tree-level mass corrections originate from the nonvanishing VEV
of the auxiliary $B$$-$$L$ $D$ field, $\left<D\right> = qg\,\xi$. 
To see this, recall that the total tree-level scalar potential in
SUGRA takes the following form,
\begin{align}
V = V_F + V_D = e^{K/M_{\rm Pl}^2} \left[
\left(W_i + \frac{W}{M_{\rm Pl}^2} K_i\right)K^{i\bar{\jmath}}
\left(\overline{W}_{\bar\jmath} + \frac{\overline{W}}{M_{\rm Pl}^2} K_{\bar\jmath}\right)
- 3\, \frac{\left|W\right|^2}{M_{\rm Pl}^2} + \frac{1}{2}\,e^{-K/M_{\rm Pl}^2}\, D^2\right] \,,
\label{eq:VFD}
\end{align}
where the indices $i$ and $\bar\jmath$ refer to differentiation w.r.t.\ to the
complex scalars $\phi_i$ and $\phi_{\bar\jmath}^*$, respectively, and where
$K^{i\bar{\jmath}}$ denotes the inverse of the K\"ahler metric,
$K^{i\bar{\jmath}} \equiv \left(K_{i\bar{\jmath}}\right)^{-1}$.
The superpotential $W$, the K\"ahler potential $K$ and the $B$$-$$L$
D-term are all nonvanishing in the true vacuum,
\begin{align}
\left<W\right> \equiv W_0 \equiv e^{-K_0/M_{\rm Pl}^2/2}\, m_{3/2}\, M_{\rm Pl}^2 \,, \quad
\left<K\right> \equiv K_0 \,, \quad \left<D\right> \equiv D_0 = qg\,\xi \,.
\label{eq:WKD0}
\end{align}
For one reason or another, these VEVs are fine-tuned such that the cosmological
constant (almost) vanishes.
This is to say that, in the low-energy vacuum, the total scalar potential is (almost) zero,
\begin{align}
\left<V\right> = 
\big<K^{i\bar{\jmath}}\mathcal{F}_i \mathcal{F}_{\bar\jmath}^*\big>
+ \frac{1}{2}\, D_0^2
- 3\, e^{K_0/M_{\rm Pl}^2}\,\frac{\left|W_0\right|^2}{M_{\rm Pl}^2} = 0 \,, \quad
\mathcal{F}_i = e^{K/M_{\rm Pl}^2/2} \left(W_i + \frac{W}{M_{\rm Pl}^2} K_i\right) \,.
\label{eq:CC0}
\end{align}
Together, Eqs.~\eqref{eq:WKD0} and \eqref{eq:CC0} allow us to solve for the gravitino
mass in terms of the total SUSY breaking scale $\Lambda_{\rm SUSY}$
(which reduces to $\mu$ in the global SUSY limit and for small $g$, see Eq.~\eqref{eq:m32}),
\begin{align}
m_{3/2}^2 = \frac{\Lambda_{\rm SUSY}^4}{3M_{\rm Pl}^2} \,, \quad
\Lambda_{\rm SUSY}^4 = F_0^2 + \frac{1}{2} D_0^2 \,, \quad
F_0^2 = \big<K^{i\bar{\jmath}}\mathcal{F}_i \mathcal{F}_{\bar\jmath}^*\big> \,.
\end{align}
Each MSSM sfermion $\tilde{f}$ now appears with a canonically normalized term in
the K\"ahler potential,
\begin{align}
K = K_0 + \tilde{f}^\dagger e^{2q_f g\,V} \tilde{f}  + \cdots 
= K_0 + \tilde{f}^\dagger \tilde{f} + \cdots \,.
\label{eq:Kff}
\end{align}
Inserting this expansion into Eq.~\eqref{eq:VFD} yields
the universal tree-level MSSM sfermion mass in PGM,
\begin{align}
V = 
\exp\left(\big|\tilde{f}\big|^2/M_{\rm Pl}^2\right) V_0 +
\left(e^{K_0/M_{\rm Pl}^2}
\frac{\left|W_0\right|^2}{M_{\rm Pl}^4}
- \frac{D_0^2}{2M_{\rm Pl}^2}\right)
\big|\tilde{f}\big|^2 + \cdots 
= V_0 + m_0^2 \,\big|\tilde{f}\big|^2 + \cdots
\end{align}
where $V_0 \equiv \left<V\right> = 0$.
Making use of the definition of the gravitino mass in
Eq.~\eqref{eq:WKD0}, we find
\begin{align}
m_0^2 = \frac{V_0}{M_{\rm Pl}^2} + e^{K_0/M_{\rm Pl}^2} \frac{\left|W_0\right|^2}{M_{\rm Pl}^4}
- \frac{D_0^2}{2M_{\rm Pl}^2} =
m_{3/2}^2 - \frac{D_0^2}{2M_{\rm Pl}^2} = m_{3/2}^2 + \Delta m_0^2 \,, \quad
\Delta m_0^2 = - \frac{q^2g^2\xi^2}{2M_{\rm Pl}^2}\,.
\label{eq:m02}
\end{align}
Here, the first contribution to $m_0$, given by the gravitino mass $m_{3/2}$,
corresponds to the universal soft mass for all sfermions in PGM \textit{in absence}
of a nonzero D-term, while the second contribution to $m_0$ represents a universal shift
in $m_0$ induced by the nonzero FI parameter $\xi$.
In the context of our SUSY breaking model and assuming that $qg\sim1$, one naively
expects $D_0 \sim \xi \sim \Lambda^2$, so that
\begin{align}
m_0 \sim m_{3/2} \sim \Delta m_0 \sim \frac{\Lambda}{M_{\rm Pl}}\,\Lambda \,.
\end{align}


This means that the $\xi$-induced shift in the soft sfermion mass,
$\Delta m_0$, may, under certain circumstances, become roughly as large as the
``bare'' soft mass in absence of a nonzero FI term, $\Delta m_0 / m_{3/2} \sim 1$.
Since the shift $\Delta m_0$ represents a tachyonic mass correction, it is, however,
important that $\Delta m_0$ never exceeds $m_{3/2}$.
This results in an upper
bound on the ratio $D_0 / F_0$,
\begin{align}
m_0^2 = \frac{1}{3M_{\rm Pl}^2}\left(F_0^2 - D_0^2\right) \geq 0
\quad\Rightarrow\quad \frac{D_0}{F_0} \leq 1 \,.
\label{eq:DF}
\end{align}
Note that this bound on the magnitude of the D-term applies independently
of the fact that the sfermion $\tilde{f}$ carries nonzero $B$$-$$L$ charge.
Instead, it holds universally for any $U(1)$ symmetry that may contribute to the
total vacuum energy with a nonvanishing D-term.
In the context of our DSB model, the VEV of the D-term is
always trivially smaller than the VEV of the IYIT F-term, at least as long as 
we stay in the weakly gauged regime, where $\gamma \ll 1$,
(see Eqs.~\eqref{eq:mum} and \eqref{eq:xiFA})
\begin{align}
\frac{D_0}{F_0} = \gamma \left[\frac{2^{1/2}}{\rho^2}\left(1-\rho^4\right)^{1/2}
+ \mathcal{O}\left(\gamma^2\right)\right] \ll 1 \,.
\end{align}
Whether or not $D_0$ always remains smaller than $F_0$ also in the strongly gauged
regime, i.e., for $\gamma \gg 1$, is an open question, which we leave for future
work.
While general SUGRA theorems suggest that this may very well be
the case~\cite{Nakayama:2007du}, it would still be interesting to determine
the precise upper bound $\left.D_0/F_0\right|_{\rm max}$ on the ratio $D_0/F_0$
in the context of the IYIT model.


Next to the universal soft mass $m_0$ in Eq.~\eqref{eq:m02}, each sfermion
receives a further tree-level mass correction $m_D$, which depends on its respective
$B$$-$$L$ charge $q_f$.
Because of the interaction with the $B$$-$$L$ $D$ field in the K\"ahler potential
(see Eq.~\eqref{eq:Kff}), each sfermion explicitly appears in $V_D$,
\begin{align}
V_D = \frac{q^2g^2}{2}\left[\xi - \frac{q_f}{q}\big|\tilde{f}\big|^2 + \cdots \right]^2
= \frac{1}{2}D_0^2 + m_D^2 \big|\tilde{f}\big|^2 + \cdots \,, \quad
m_D^2 = -q_f g\, D_0 = - q\,q_f g^2\xi \,,
\end{align}
so that we eventually obtain for the total tree-level mass $m_{\tilde{f}}^{\rm tree}$
of an MSSM sfermion,
\begin{align}
\left(m_{\tilde{f}}^{\rm tree}\right)^2 = m_0^2 + m_D^2\,, \quad
m_0^2 = m_{3/2}^2 - \frac{q^2g^2\xi^2}{2M_{\rm Pl}^2} \,, \quad
m_D^2 = - q\,q_f g^2\xi \,.
\label{eq:mf2}
\end{align}
We hence see that sfermions $\tilde{f}$ with charge $q_f$ such
that $q_f\xi >0$ acquire a negative mass squared as long as the FI parameter
$\xi$ is not substantially suppressed w.r.t.\ the dynamical scale $\Lambda$,
\begin{align}
m_0 \sim \frac{\Lambda^2}{M_{\rm Pl}} \,, \quad \xi \sim \Lambda^2 
\,, \quad q\,\xi >0  \quad\Rightarrow\quad
\left(m_{\tilde{f}}^{\rm tree}\right)^2 \sim - \Lambda^2 \left[1 +
\mathcal{O}\left(\frac{\Lambda^2}{M_{\rm Pl}^2}\right)\right] \,.
\label{eq:m0estimate}
\end{align}
This poses a serious problem, which, in general, may be regarded as a fundamental obstacle
to identifying any gauged $U(1)$ flavor symmetry featuring a nonzero D-term with
$B$$-$$L$.
Of course, a trivial way out of this problem is
to assume an extremely small $B$$-$$L$ gauge coupling, $g \lesssim \Lambda/M_{\rm Pl}$,
so as to suppress $m_D^2$ by a factor $g^2 \lesssim \left(\Lambda/M_{\rm Pl}\right)^2$.
In the case of PGM, where one typically has
$\Lambda \sim 10^{12}\,\textrm{GeV}$ (see Eq.~\eqref{eq:Lambdaeta}),
this would mean that $g$ should take at most a value of $\mathcal{O}\left(10^{-6}\right)$.
Such a tiny gauge coupling is certainly rather unusual, which leads one
to wonder whether there is not a possibility to somehow lift the upper bound on $g$
by means of another suppression mechanism.


\subsection[Suppressing the $B$$-$$L$ D-term by means of an approximate flavor symmetry]
{Suppressing the \boldmath{$B$$-$$L$} D-term by means of an approximate flavor symmetry}


One of the main conceptual achievements in the present paper is the
realization that this is indeed possible!
Our main observation is that the FI parameter $\xi$ itself could be
parametrically suppressed, $\xi/\Lambda^2 \lesssim \left(\Lambda/M_{\rm Pl}\right)^2$,
in consequence of an enhanced flavor symmetry.
The logic behind this idea is the following:
Under generic circumstances, all we could say about $\xi$ is
that it arises from a combination of scalar VEVs
in the D-term scalar potential.
If, e.g., two scalar fields $\phi_\pm$ with charges $\pm 1$ were involved
in the generation of $\xi$, we would write
\begin{align}
V_D = \frac{g^2}{2} \left[\big<\left|\phi_-\right|^2\big> - 
\big<\left|\phi_+\right|^2\big> - q_f
\big|\tilde{f}\big|^2 + \cdots \right]^2 \,, \quad
\xi = \big<\left|\phi_-\right|^2\big> - \big<\left|\phi_+\right|^2\big> \,.
\end{align}
At this level of the description, a suppressed value of $\xi$ would merely correspond
to a fine-tuning among the VEVs of $\phi_+$ and $\phi_-$, which might appear
very unnatural at first sight.
In order to explain why $\xi$ should be much smaller than one would naively
expect, $\left|\xi\right| \ll \big<\left|\phi_\pm\right|^2\big>$, we therefore
require a more detailed description of how $\xi$ is actually generated in the course
of spontaneous SUSY breaking---which is exactly the case in the DSB model studied in
the present paper.
Within the IYIT model supplemented by a weakly gauged flavor symmetry,
we are able to derive an explicit expression for $\xi$ in terms of the
underlying model parameters (see Eq.~\eqref{eq:xiFA}).
The question as to whether or not $\xi$ has a chance of ending up suppressed is
then no longer a question pertaining to scalar VEVs, but rather to the
Yukawa couplings in the Lagrangian.
This opens up the possibility to render $\xi$ arbitrarily small by imposing an
approximate flavor symmetry among these couplings.


Recall that the parameter $\rho$ in Eq.~\eqref{eq:xiFA} characterizes the quality
of the ``exchange symmetry'' ``$+$'' $\leftrightarrow$ ``$-$'' in the charged
meson sector (see Eq.~\eqref{eq:lrg} and Footnote~\ref{fn:rho}).
In the limit of an exact exchange symmetry, $\rho$ goes to $1$ and the FI parameter
$\xi$ trivially vanishes altogether,
\begin{align}
\xi = \frac{2\left(1-\rho^4\right)^{1/2}}{\rho^2}\left(\frac{\Lambda}{\eta}\right)^2
\quad\overset{\rho\rightarrow1}{\longrightarrow}\quad 0 \,.
\end{align}
However, before we put too much trust in this limit, we first have to clarify the actual
meaning of this exchange symmetry.
To do so, note that the exchange symmetry
``$+$'' $\leftrightarrow$ ``$-$'' can, in fact, be re-formulated as a $Z_2$
parity acting on the following linear combinations of the fields $M_\pm$ and $Z_\pm$,
\begin{align}
\frac{1}{\sqrt{2}}\left(M_+ + M_-\right) \,, \quad
\frac{1}{\sqrt{2}}\left(Z_+ + Z_-\right) \,, \quad
\frac{1}{\sqrt{2}}\left(M_+ - M_-\right) \,, \quad
\frac{1}{\sqrt{2}}\left(Z_+ - Z_-\right) \,,
\end{align}
where the first two linear combinations transform even and the last
two linear combinations transform odd under this $Z_2$ parity. 
For generic Yukawa couplings $\lambda_0^a$ (see Eq.~\eqref{eq:Weff}),
this $Z_2$ parity can, however, \textit{not} be realized at the level
of the fundamental quark fields above the dynamical scale $\Lambda$. 
For instance, if we tried to realize the $Z_2$ exchange symmetry
by assigning the following transformation behavior to the four fundamental
quark fields,
\begin{align}
\Psi^1 \leftrightarrow \Psi^3 \,, \quad \Psi^2 \leftrightarrow \Psi^4 \,, 
\end{align}
the gauge-invariant composite meson fields at low energies,
$M_+ \propto \left<\Psi^1\Psi^2\right>$,
$M_- \propto \left<\Psi^3\Psi^4\right>$,
$M_0^1 \propto \left<\Psi^1\Psi^3\right>$,
$M_0^2 \propto \left<\Psi^1\Psi^4\right>$,
$M_0^3 \propto \left<\Psi^2\Psi^3\right>$,
$M_0^4 \propto \left<\Psi^2\Psi^4\right>$,
would transform as follows,
\begin{align}
M_+ \leftrightarrow M_- \,, \quad
M_0^1 \leftrightarrow M_0^1 \,, \quad
M_0^2 \leftrightarrow M_0^3 \,, \quad
M_0^4 \leftrightarrow M_0^4 \,.
\end{align}
In this case, it would not be sufficient to simply set $\lambda_+  = \lambda_-$
in order to realize the exchange symmetry in the superpotential; we also would
have to require that $\lambda_0^2  = \lambda_0^3$.
This tells us that it is, in general, not possible to identify the $Z_2$ symmetry
as a subgroup of the global $SU(4)$ flavor symmetry, which we obtain in the limit
of equal Yukawa couplings, $\lambda_{ij} \equiv \lambda$.
For generic Yukawa couplings in the neutral meson sector, the exchange symmetry
in the charged meson sector should rather be regarded as an accidental symmetry
of the low-energy effective theory, which we happen to encounter
once we set $\lambda_+  = \lambda_-$.
As nothing but an accidental symmetry of the effective superpotential,
the exchange symmetry is then expected to be explicitly broken by higher-order terms in
the effective K\"ahler potential, so that we basically loose all control over its quality.


The lesson from these considerations is that it is not enough to simply
send the parameter $\rho$ to $1$ in order to suppress the FI parameter $\xi$. 
Instead, we have to impose a larger (approximate) global flavor symmetry, not merely
a $Z_2$ exchange symmetry in the charged meson sector.
Here, an obvious choice is to require the full $SU(4)$ flavor symmetry to be
approximately realized in the IYIT sector, so that $Z_2 \subset SU(4)$.
In this case, it is then possible to identify the $Z_2$ exchange symmetry with
a global flavor symmetry of the fundamental theory at high energies and it
is conceivable that the parameter $\rho$ indeed takes a value very close to $1$.
Meanwhile, we caution that the $SU(4)$ symmetry of the IYIT
superpotential should not attain an arbitrarily good quality,
as this would render the three low-energy vacua of the
IYIT model degenerate (see Eq.~\eqref{eq:V123} and Footnote~\ref{fn:flavsymm}).
In fact, in the limit of an exact $SU(4)$ symmetry, the vacua I, II and III
become connected to each other via four flat directions that may
be regarded as coordinates of the compact space $SO(6)/SO(5)$~\cite{Chacko:1998si},
and which might cause serious problems at low energies.%
\footnote{The compact space $SO(6)/SO(5)$ is, in total, $5$-dimensional.
Recall, however, that one flat direction corresponds to the $B$$-$$L$ goldstone phase
that is absorbed by the $B$$-$$L$ vector boson upon spontaneous symmetry breaking.}
On the other hand, as long as the $SU(4)$ symmetry is only approximately realized,
these four directions in field space have masses that scale with the 
differences between the geometric means
$\lambda=\left(\lambda_+\lambda_-\right)^{1/2}$,
$\lambda_{14}=\left(\lambda_0^1\lambda_0^4\right)^{1/2}$ and 
$\lambda_{23}=\left(\lambda_0^2\lambda_0^3\right)^{1/2}$.
For $\lambda_0^1 = \lambda_0^4$ and $\lambda_0^2 = \lambda_0^3$, e.g.,
we find that the neutral mesons $M_0^a$ give rise to two complex mass
eigenstates, $m_{14}^-$ and $m_{23}^-$,
with almost vanishing masses~\cite{Harigaya:2015soa},
\begin{align}
m_{m_{14}^-}^2 = \left(\lambda_{14}^2 - \lambda^2\right)\left(\frac{\Lambda}{\eta}\right)^2
\,,\quad
m_{m_{23}^-}^2 = \left(\lambda_{23}^2 - \lambda^2\right)\left(\frac{\Lambda}{\eta}\right)^2
\,.
\end{align}
Requiring that the masses squared of these complex scalars remain positive,
$\lambda < \min\left\{\lambda_{14},\lambda_{23}\right\}$, is then equivalent
to the condition that the $B$$-$$L$-breaking vacuum should be the lowest-lying among
the three low-energy vacua of the IYIT sector (see the discussion below Eq.~\eqref{eq:V123}).


Another caveat applying to the quality of the global flavor symmetry in the
IYIT sector pertains to the anomaly-free $Z_4$ symmetry
which is realized even for all Yukawa couplings $\lambda_{ij}$
being different (see Eq.~\eqref{eq:breaking}).
This symmetry is broken by the VEVs of the charged meson fields, together with SUSY
and $B$$-$$L$, down to a $Z_2$ parity (under which all quarks transform
odd, $\Psi^i \rightarrow - \Psi^i$).
If this symmetry was exact, its spontaneous breaking would result in the formation
of \textit{stable} domain walls, which might have disastrous cosmological 
consequences~\cite{Zeldovich:1974uw}.%
\footnote{Whether or not stable $Z_4$ domain walls would lead to cosmological
problems depends on the scale of inflation:
If the Hubble scale during inflation, $H_0$, is low, $H_0 \lesssim \Lambda$, the
$Z_4$-breaking phase transition takes place during inflation and all dangerous
domain walls are inflated away.
However, in the case of large-scale inflation, $H_0 \gtrsim \Lambda$, the $Z_4$
symmetry is only broken after the end of inflation, so that the associated
formation of domain walls would pose a problem.
Note that similar considerations may also help to explain why the SUSY breaking
scale needs, in fact, to be much higher than one would naively expect according to arguments
based on the idea of electroweak naturalness~\cite{Harigaya:2015yla}.}
Thus, also the $Z_4$ symmetry of the IYIT superpotential should
only be approximately realized, so that its breaking leads at most to the formation
of \textit{unstable} domain walls, which quickly annihilate after their production.
This is, e.g., achieved by higher-order terms in the
K\"ahler potential that explicitly break $Z_4$.


After these qualitative remarks, we are now ready to study the suppression of the FI
parameter $\xi$ in more quantitative terms. 
To do so, let us first expand $m_{\tilde{f}}^{\rm tree}$ in Eq.~\eqref{eq:mf2} around $\rho = 1$,
\begin{align}
\left(m_{\tilde{f}}^{\rm tree}\right)^2 = m_{3/2}^2 \left[1 - 6^{1/2}\lambda\,\gamma^2\,\frac{q_f}{q}
\frac{M_{\rm Pl}}{m_{3/2}}\,\epsilon + \mathcal{O}\left(\epsilon^2\right)\right] \,, \quad
\epsilon \equiv \left(1-\rho^4\right)^{1/2} \,.
\label{eq:mf2eps}
\end{align}
Here, we have introduced the parameter $\epsilon \in [0,1]$ to describe
small deviations from the flavor-symmetric limit, $\epsilon \rightarrow 0$.
Note that $\epsilon$ not only directly parametrizes the suppression of
$\xi$, it also corresponds to the relative difference between
the Yukawa couplings $\lambda_+$ and $\lambda_-$ squared,
\begin{align}
\xi = \frac{2\,\epsilon}{\left(1-\epsilon^2\right)^{1/2}}
\left(\frac{\Lambda}{\eta}\right)^2 \,, \quad
\epsilon = \left|\frac{\lambda_+^2 - \lambda_-^2}{\lambda_+^2 + \lambda_-^2}\right| \,.
\end{align}
Our philosophy in the following will now be that the parameter $\epsilon$ can, in principle,
take arbitrarily small values, so that the exchange symmetry in the charged
meson sector becomes arbitrarily good. 
We emphasize that this is not in contradiction with our above remarks
regarding the quality of the $SU(4)$ or $Z_4$ flavor symmetries, as it only
pertains to the relation between the Yukawa couplings $\lambda_+$ and $\lambda_-$.
We can always render the total flavor symmetry sufficiently broken
by retaining a (small) hierarchy among $\lambda$, $\lambda_{14}$
and $\lambda_{23}$, irrespectively of how close $\lambda_+$ and $\lambda_-$
are to each other.
Given the sfermion mass $m_{\tilde{f}}^{\rm tree}$ in Eq.~\eqref{eq:mf2eps},
we then find that, in the small-$\epsilon$ regime, the tree-level bound on
the gauge coupling constant $g$ scales as follows with the suppression factor $\epsilon$,
\begin{align}
m_{\tilde{f}}^{\rm tree} \geq 0  \quad\Rightarrow\quad g \leq g_{\rm max}^{\rm tree} \approx
\left(\frac{\lambda}{6^{1/2}\,q\,q_f}
\frac{m_{3/2}}{M_{\rm Pl}}\frac{1}{\epsilon}\right)^{1/2} \,.
\end{align}
For $q=2$ and $q_f = 1$,
we can, hence, lift the bound on $g$ to some $\mathcal{O}(1)$ value, if
$\epsilon$ is of $\mathcal{O}\left(10^{-13}\right)$,
\begin{align}
g_{\rm max}^{\rm tree} \simeq 0.9 \, \bigg(\frac{\lambda}{1}\bigg)^{1/2}
\bigg(\frac{m_{3/2}}{1000\,\textrm{TeV}}\bigg)^{1/2}
\left(\frac{10^{-13}}{\epsilon}\right)^{1/2} \,,
\label{eq:gmaxtree}
\end{align}
so that the magnitude of the FI parameter $\xi$ is pushed just below the
gravitino mass squared,
\begin{align}
\xi \simeq  \frac{m_{3/2}^2}{q\,q_f\left(g_{\rm max}^{\rm tree}\right)^2}
\simeq 0.6\, m_{3/2}^2 \,.
\end{align}


We, thus, find that an approximate flavor symmetry among the
couplings of the IYIT sector allows us to sufficiently suppress the $B$$-$$L$
D-term.
Here, the key feature of our analysis has been the calculability of the D-term
in the context of the IYIT model, due to which we were able to compute an explicit
expression for $\xi$ in terms of the underlying model parameters
(see Eq.~\eqref{eq:xiFA}).
We believe that this feature of the IYIT model readily generalizes to a variety of
other DSB models.
This means that a number of D-terms (belonging to certain gauged flavor symmetries),
which might appear very large at first sight, may actually turn out to be substantially
suppressed, as long as one imposes the right flavor symmetry on the SUSY-breaking
dynamics.
While, in retrospective, this result may appear trivial,
we emphasize the importance of having concrete examples at one's disposal
that illustrate, within the context of specific models, how dynamically
generated D-terms may indeed be suppressed by means of approximate flavor symmetries.
For this reason, one of the main motivations behind the present paper is to provide
just such an example.


\subsection{Gauge-mediated contributions to the MSSM sfermion masses}


This is, however, not the end of the story.
So far, we have only considered
the tree-level corrections to the masses of the MSSM sfermions.
Besides that, we also have to take into account that the nonzero charges of the
SUSY-breaking fields $Z_\pm$ result in a mass splitting within the $B$$-$$L$
vector multiplet. 
The $B$$-$$L$ gauge DOFs thus act as gauge messengers
that induce gauge-mediated sfermion masses at the loop level.
Here, the most important (one-loop) correction is given as~\cite{Intriligator:2010be}
\begin{align}
\left(m_{\tilde{f}}^{\rm 1-loop}\right)^2 = - \frac{q_f^2g^2}{32\pi^2}\, m_V^2\,
\ln\left[\frac{m_{\tilde{a}}^8}{m_V^6 m_\phi^2}\right] =
- \frac{q_f^2g^2}{32\pi^2}\, m_V^2\,
\ln\left[\frac{\left(m_V^2+m^2\right)^4}{m_V^6\left(m_V^2 + 2m^2\right)}\right] \,,
\label{eq:mf2loop}
\end{align}
with $m_V$, $m_{\tilde{a}}$ and $m_\phi$ denoting the masses of the vector
boson $A_\mu$, gaugino $\tilde{a}$ and real scalar $\phi$ contained in the ``massive
$B$$-$$L$ vector multiplet'' $V_A$, respectively,
(see Eqs.~\eqref{eq:mum}, \eqref{eq:KVA} and \eqref{eq:ximV})%
\footnote{More precisely, $\tilde{a}$ is the fermionic component of the
$B$$-$$L$ goldstone multiplet
$A$, which shares a Dirac mass term with the fermionic component of the linear
combination $Y$ (see Eq.~\eqref{eq:WXYA}), and $\phi$ is the real part of the
complex scalar contained in the goldstone multiplet.
$\tilde{a}$ and $\phi$ therefore correspond to the fermionic and scalar partners of
the goldstone phase $a$~\cite{Domcke:2014zqa}.
Similarly, if $A$ was to be identified with the chiral axion multiplet in a
supersymmetric implementation of the PQ mechanism, $\tilde{a}$ and $\phi$
would be referred to as the axino and the saxion, respectively~\cite{Harigaya:2015soa}.}
\begin{align}
m_V^2 = 2\, q^2 g^2 f_A^2 \,, \quad m_{\tilde{a}}^2 = m_V^2 + m^2 \,, \quad
m_\phi^2 = m_V^2 + 2\,m^2 \,, \quad
m^2 = \rho^2 \lambda^2 \left(\frac{\Lambda}{\eta}\right)^2 \,.
\end{align}
The effective one-loop correction in Eq.~\eqref{eq:mf2loop} contributes
to the total sfermion mass at $\mathcal{O}\left(\gamma^4\right)$,
\begin{align}
\left(m_{\tilde{f}}^{\rm 1-loop}\right)^2 = \left(\frac{3}{2}\right)^{1/2}
\frac{\gamma^4\lambda^3}{8\pi^2} \left(\frac{q_f}{q}\right)^2 \frac{M_{\rm Pl}}{m_{3/2}}
\left[\ln 128 + 6 \ln \gamma + \mathcal{O}\left(\gamma^2\right)\right] m_{3/2}^2 \,,
\end{align}
which is always negative.
That is, even when the tree-level, $\xi$-induced contribution to the MSSM sfermion masses
is sufficiently suppressed, the gauge-mediated one-loop contribution in Eq.~\eqref{eq:mf2loop}
may still render the MSSM sfermions tachyonic.
To prevent this from happening, the gauge coupling constant $g$ must remain small enough,
so that $\big|m_{\tilde{f}}^{\rm 1-loop}\big|$ is always smaller than $m_{3/2}$.


\begin{figure}[t]
\centering
\includegraphics[width=0.67\textwidth]{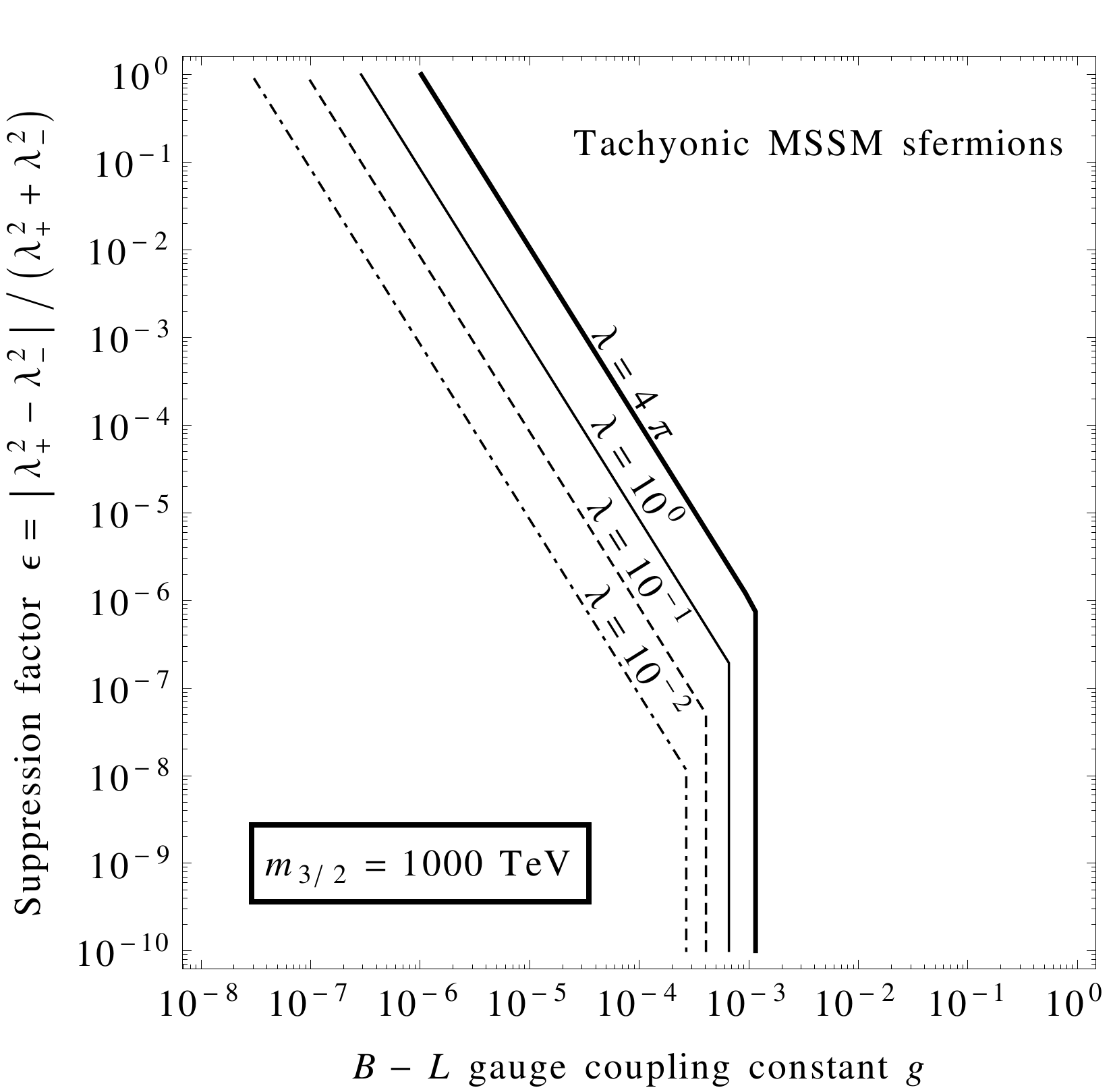}
\caption{Bound on the $B$$-$$L$ gauge coupling constant $g$ as a function of
the suppression factor $\epsilon$ (see Eq.~\eqref{eq:mf2eps}) and the IYIT Yukawa
coupling $\lambda$ (see Eq.~\eqref{eq:lrg}) for $m_{3/2} = 1000\,\textrm{TeV}$.
For very small values of $\epsilon$, the tree-level, $\xi$-induced contributions to
the MSSM sfermion masses are negligible and $g$ is constrained according to the
loop-level bound in Eq.~\eqref{eq:gmax}.
For larger values of $\epsilon$, the tree-level bound in Eq.~\eqref{eq:gmaxtree}
then becomes more stringent than the one in Eq.~\eqref{eq:gmax}, so that
$g$ becomes even more strongly constrained.
Note that $\lambda$ should not be chosen much smaller than
$\mathcal{O}\left(10^{-3}\right)$, since otherwise the VEV of the
SUSY-breaking field $X$ in SUGRA, $\left<X\right> \sim 16\pi^2 / \lambda^3\,m_{3/2}$
would begin to exceed the Planck scale~\cite{Domcke:2014zqa,Harigaya:2015soa}.
At the same time, unitarity restricts $\lambda$ to take at most a value of
$\lambda_{\rm max} \simeq \eta \simeq 4\pi$.}
\label{fig:gbounds}
\end{figure}


This results in an absolute upper bound on the gauge coupling that cannot be
lifted any further, even if we tune the suppression factor $\epsilon$
in Eq.~\eqref{eq:gmaxtree} to an arbitrarily small value,
\begin{align}
\left|m_{\tilde{f}}^{\rm 1-loop}\right| \leq m_{3/2} \quad\Rightarrow\quad
g \leq g_{\rm max}^{\rm loop} = \frac{\lambda}{2^{7/6}\,q}
\exp\left(\frac{1}{4}\,W_{-1}\left[-\frac{2^{1/6}\,512\pi^2}{3^{3/2}\lambda^3}
\left(\frac{q}{q_f}\right)^2 \frac{m_{3/2}}{M_{\rm Pl}}\right]\right) \,,
\label{eq:gmax}
\end{align}
where $W_{-1}$ denotes the lower branch of the Lambert $W$ function or product logarithm
(which can take values $-\infty \leq W_{-1} \leq -1$ and which satisfies
$x = W_{-1}(x) e^{W_{-1}(x)}$, so that $W_{-1}\left(x\,e^x\right) = x$).
For $m_{3/2} = 1000\,\textrm{TeV}$ and $\lambda = 1$, the bound $g_{\rm max}^{\rm loop}$
evaluates to $g_{\rm max}^{\rm loop} \sim 10^{-3}$, which means that a
gauge coupling constant of $\mathcal{O}(1)$ is, in fact, unviable in our set-up.
On the other hand, it is worth noting that, by imposing an approximate flavour
symmetry, we were able to relax the naive bound on $g$ resulting from the tree-level
D-term scalar potential, $g \lesssim 10^{-6}$ (see our discussion below
Eq.~\eqref{eq:m0estimate}), by three orders of magnitude, which is
a remarkable improvement.
Finally, we note that, depending on the value of the suppression factor
$\epsilon$, either the tree-level bound on $g$ in Eq.~\eqref{eq:gmaxtree}
or the loop-level bound in Eq.~\eqref{eq:gmax} dominates.
This is summarized in Fig.~\ref{fig:gbounds}, where we show the maximally allowed value
of $g$ as a function of $\epsilon$ for different values of $\lambda$.
 

\section{Conclusions and outlook}
\label{sec:conclusions}


The IYIT model is an instructive and easy-to-handle toy model for examining
how the dynamics of dynamical SUSY breaking might be related to
other beyond-the-standard-model phenomena. 
In particular, the global $U(1)_A$ flavor symmetry present in the IYIT
tree-level superpotential is well suited to be identified with other
commonly studied local or global $U(1)$ symmetries:
(i) In~\cite{Domcke:2014zqa}, e.g., this $U(1)_A$ symmetry has been promoted to
a weakly gauged FI symmetry, $U(1)_A \rightarrow U(1)_{\rm FI}$, in order to demonstrate
how dynamical SUSY breaking may entail the generation of a field-dependent FI-term
in field theory.
(ii) Meanwhile, in~\cite{Harigaya:2015soa}, the same $U(1)_A$ symmetry has been identified with
the global PQ symmetry, $U(1)_A \rightarrow U(1)_{\rm PQ}$, in order to point out a
possibility how the dynamical breaking of SUSY may also give rise
to a QCD axion that is capable of solving the strong $CP$ problem.
(iii) And in the present paper, we have finally promoted the $U(1)_A$
symmetry to a weakly gauged $B$$-$$L$ symmetry, $U(1)_A \rightarrow U(1)_{B-L}$,
in order to illustrate how the paradigm of pure gravity
mediation (PGM) may be implemented into concrete models of dynamical SUSY breaking.


This has led us to a number of interesting conceptual and phenomenological observations.
For one thing, we have described a mechanism by means of which one is able to
sufficiently suppress the $B$$-$$L$ D-term, so that it no longer poses a threat
to low-energy phenomenology:
In the context of the IYIT model, we were able to derive an explicit expression for the
$B$$-$$L$ FI parameter $\xi$ in terms of the Yukawa couplings appearing in the IYIT
superpotential.
We then found that, by imposing an approximate flavor symmetry on the SUSY-breaking
dynamics, the magnitude of the D-term in the $B$$-$$L$ gauge sector can be rendered
arbitrarily small.
We are confident that similar results also hold for D-terms associated with other
gauged flavor symmetries in the context of other DSB models.
For another thing, we have identified a direct relation between the heavy neutrino mass scale
in the seesaw extension of the MSSM, $\Lambda_N$, and the gravitino mass $m_{3/2}$:
If the spontaneous breakings of SUSY and $B$$-$$L$ should really be tied to each other
similarly as in the set-up investigated in this paper, we expect that
$\Lambda_N \sim m_{3/2}$.
The heavy neutrino mass scale then ends up being much smaller than naively
expected, i.e., much smaller than the GUT scale,
$\Lambda_N \ll \Lambda_{\rm GUT} \sim 10^{16}\,\textrm{GeV}$, which has
profound implications for cosmology.


For heavy Majorana neutrinos as ``light'' as $m_{3/2} \sim 1000\,\textrm{TeV}$,
we are, e.g., no longer able to rely on standard thermal leptogenesis to account
for the origin of the baryon asymmetry in the universe.
Instead, leptogenesis should proceed at a much lower energy scale, like in the
case of resonant leptogenesis or nonthermal leptogenesis via
inflaton decay~\cite{Lazarides:1991wu}.
Here, we note that such a scenario fits together particularly well with
the notion of thermal wino dark matter in the framework of PGM.
As has recently been shown, a \textit{thermal} relic abundance of MSSM winos with a mass
around $3\,\textrm{TeV}$ allows to nicely reproduce the antiproton-to-proton ratio
measured by the AMS-02 experiment in cosmic rays~\cite{Ibe:2014qya}.
Therefore, in order to avoid overproduction of \textit{nonthermal} winos in gravitino decays
after reheating, the reheating temperature after inflation should not be too high.
This favors some form of low-scale leptogenesis over standard thermal leptogenesis, which
agrees with the fact that our model predicts a low neutrino mass scale $\Lambda_N$.
In addition to that, in the particular case of nonthermal leptogenesis via inflaton decay,
the reheating temperature should also not be too low, $T_{\rm rh} \gtrsim 10^6\,\textrm{GeV}$,
since otherwise leptogenesis fails to generate a sufficient baryon asymmetry.
In this case, the heavy Majorana neutrinos must then have a mass of at least
$\mathcal{O}\left(1000\right)\,\textrm{TeV}$, which, in the context of our model,
translates into $m_{3/2} \gtrsim 1000\,\textrm{TeV}$---nicely
in accord with the general idea behind the concept of PGM.
Under the specific assumption of nonthermal leptogenesis, the connection between $\Lambda_N$
and $m_{3/2}$ discussed in this paper therefore automatically entails a possible answer to
the fundamental question as to why SUSY apparently needs to be broken
at a scale that is much higher than naively expected according to electroweak naturalness
(i.e., as to why $m_{3/2} \gg 100\,\textrm{GeV}$, so that we have not yet seen SUSY at colliders).
This is an intriguing observation, which directly follows from the connection between
the spontaneous breakings of SUSY and $B$$-$$L$ proposed in this paper
(see also \cite{Harigaya:2015yla} for a similar argument).


Another prediction of our model is the fact that the $B$$-$$L$ gauge coupling
constant can at most be as large as $\mathcal{O}\left(10^{-3}\right)$.
For larger values of $g$, the SUSY-breaking mass splitting within
the massive $B$$-$$L$ vector multiplet results in too large (negative) gauge-mediated
contributions to the MSSM sfermion masses.
This upper bound on $g$ justifies, \textit{a posteriori}, that we have performed all of
our calculations in the weakly gauged limit.
From a theoretical point of view, it would, however, still be interesting to
generalize our analysis to arbitrary values of the gauge coupling constant.
We anticipate such a study to lead to conceptual insights,
which may very well imply more general applications for dynamical SUSY breaking
and/or gauge mediation than our study for the special case of a local $B$$-$$L$ symmetry.
Moreover, such an analysis would allow to determine the global maximum of
the ratio $D_0/F_0$ in the IYIT model (see Eq.~\eqref{eq:DF}),
which would also be of great theoretical interest.
Last but not least, we point out that, if the bounds on $\xi$ and $g$
derived in this paper should only be \textit{marginally} satisfied, we would expect
a characteristic modulation of the MSSM sparticle spectrum compared to the
``pure PGM'' case which is determined by the $B$$-$$L$ charges of the MSSM
sfermions.
This could, in particular, result in a sizable mass gap between light sleptons 
and heavy squarks---an intriguing possibility, which deserves further study as well.


\subsubsection*{Acknowledgements}

The authors would like to thank K.~Harigaya and M.~Ibe for helpful
discussions and comments.
K.\,S.\,B.\ is grateful to Kavli IPMU at the University of Tokyo for hospitality
during the early stages of this project.
K.\,S.\ wishes to thank the Institute of Theoretical Physics at Peking University
for hospitality during the late stages of this project.
This work has been supported in part by the U.\,S.\ Department of Energy Grant
No.\ DE-SC0010108 (K.\,S.\,B.),
Grants-in-Aid for Scientific Research from the Ministry of Education, Culture, Sports,
Science and Technology (MEXT), Japan, No.\ 26104009 and No.\ 26287039 (T.\,T.\,Y.)
as well as by the World Premier International Research Center Initiative
(WPI), MEXT, Japan (K.\,S.\ and T.\,T.\,Y.).



\end{document}